# Oversampling smoothness (OSS): an effective algorithm for phase retrieval of noisy diffraction intensities

**Jose A Rodriguez**[a,b,d], **Rui Xu**[c,d], **Chien-Chun Chen**[c], **Yunfei Zou**[c], **Jianwei Miao**[c*]

[a]Department of Chemistry and Biochemistry, UCLA-DOE Institute for Genomics and Proteomics, University of California, Los Angeles, California, 90095, USA.
[b]Howard Hughes Medical Institute (HHMI).
[c]Department of Physics & Astronomy and California NanoSystems Institute, University of California, Los Angeles, California, 90095, USA.
[d]These authors contributed equally to this work.
*Correspondence email: miao@physics.ucla.edu.

**Synopsis**

An iterative phase retrieval algorithm, termed oversampling smoothness (OSS), has been developed to reconstruct fine features in weakly scattered objects such as biological specimens from noisy experimental data. OSS is expected to find application in the rapidly growing coherent diffraction imaging field as well as other disciplines where phase retrieval from noisy Fourier magnitudes is needed.

**Abstract**

Coherent diffraction imaging (CDI) is high-resolution lensless microscopy that has been applied to image a wide range of specimens using synchrotron radiation, X-ray free electron lasers, high harmonic generation, soft X-ray laser and electrons. Despite these rapid advances, it remains a challenge to reconstruct fine features in weakly scattering objects such as biological specimens from noisy data. Here we present an effective iterative algorithm, termed oversampling smoothness (OSS), for phase retrieval of noisy diffraction intensities. OSS exploits the correlation information among the pixels or voxels in the region outside of a support in real space. By properly applying spatial frequency filters to the pixels or voxels outside the support at different stage of the iterative process (i.e. a smoothness constraint), OSS finds a balance between the hybrid input-output (HIO) and error reduction (ER) algorithms to search for a global minimum in solution space, while reducing the oscillations in the reconstruction. Both our numerical simulations with Poisson noise and experimental data from a biological cell indicate that OSS consistently outperforms the HIO, ER-HIO and noise robust (NR)-HIO algorithms at all noise levels in terms of accuracy and consistency of the reconstructions. We expect OSS to find application in the rapidly growing CDI field as well as other disciplines where phase retrieval from noisy Fourier magnitudes is needed[1].

## 1. Introduction

When a coherent wave illuminates a non-crystalline specimen or a nanocrystal, the diffraction intensities in the far field are continuous and can be sampled at a frequency finer than the Nyquist interval (i.e. oversampled). If the sampling frequency is sufficiently fine such that the number of independent equations of the intensities is greater than or equal to the number of unknown variables describing the sample structure, the phase information is, in principle, encoded inside the diffraction intensities (Sayre, 1952; Miao *et al.*, 1998), and can be directly retrieved by using an iterative algorithm (Fienup, 1978; Fienup, 1982; Marchesini *et al.*, 2003; Elser, 2003; Luke, 2005; Marchesini, 2007; Chen *et al.*, 2007). Since the first experimental demonstration of this lensless imaging technique by Miao et al. (1999), CDI has undergone rapid development using synchrotron radiation (Robinson and Harder, 2009; Chapman and Nugent, 2010; Miao *et al.*, 2012; Miao *et al.*, 2002; Pfeifer *et al.*, 2006; Chapman *et al.*, 2006), X-ray free electron lasers (X-FELs) (Chapman *et al.*, 2006b; Seibert *et al.*, 2011; Mancuso *et al.*, 2010; Schlichting and Miao, 2012), high harmonic generation (Sandberg *et al.*, 2007; Sandberg *et al.*, 2008; Ravasio *et al.*, 2009; Seaberg *et al.*, 2011), soft X-ray lasers (Sandberg *et al.*, 2007) and electrons (Zuo *et al.*, 2003; Dronyak *et al.*, 2009; De Caro 2010). Various forms of CDI methods have been developed, including plane-wave CDI (Miao *et al*., 1999; Miao *et al.*, 2002; Chapman *et al.*, 2006; Chapman *et al.*, 2006b; Seibert *et al.*, 2011; Sandberg *et al.*, 2007; Zuo *et al.*, 2003), Bragg CDI (Robinson and Harder, 2009; Pfeifer *et al.*, 2006; Newton *et al.*, 2010), scanning (or ptychographic) CDI (Rodenburg *et al*., 2007; Thibault *et al.*, 2008; Giewekemeyer *et al.*, 2010), reflection CDI ( Marathe *et al.*, 2010; Roy *et al.*, 2011), Fresnel CDI (Williams *et al.*, 2006), and others (Abbey *et al.*, 2008; Szameit et al. 2012). Although significant advances have been made over the past few years to develop CDI methods and pursue their application in materials science, nanoscience and biology, it remains a challenge to reconstruct fine features in weakly scattering objects such as biological specimens from noisy experimental data.

Overcoming this challenge requires i) construction of dedicated CDI instruments for measuring high quality diffraction patterns and ii) development of more advanced algorithms for phase retrieval of noisy data. In this article, we will focus on the latter. To date, a number of iterative algorithms have been developed to recover the phase information from oversampled diffraction patterns (Fienup, 1978; Fienup, 1982; Marchesini *et al.*, 2003; Elser, 2003; Luke, 2005; Marchesini, 2007; Chen *et al.*, 2007). The most widely used is arguably the HIO algorithm (Fienup, 1982). HIO iterates back and forth between real and reciprocal space. In real space the no-density region and non-negativity of the electron density are used as constraints, and in reciprocal space the Fourier magnitudes as constraints. An important feature of HIO is its ability to avoid local minima and converge to a global minimum for noise-free diffraction patterns. However, when a diffraction pattern is corrupted by experimental noise, the real space image usually oscillates as a function of the number of iterations. In practice, the ER algorithm can be combined with HIO to improve the performance, but this approach suffers from stagnation and can be trapped in local minima of solution space (Fienup, 1982; Pfeifer *et al.*, 2006). In 2010, a new real space constraint (i.e. smoothness) was first developed by exploiting the correlation information among pixels or voxels in the no-density region outside a support (Raines *et al.*, 2010). Most recently the NR framework was introduced with the intent of improving the performance of existing phase retrieval algorithms such as HIO, in the face of experimental noise (Martin *et al.*, 2012). In this article, we fully exploit the smoothness constraint in real space and develop an effective iterative algorithm (OSS) for phase retrieval of noisy diffraction patterns. Both our numerical simulation and experimental results have demonstrated that OSS consistently outperforms HIO, ER-HIO and NR-HIO for the reconstruction of weakly scattering objects such as biological specimens at all noise levels.

---
[1] The Matlab source code of the OSS algorithm will be posted on a public website (www.physics.ucla.edu/research/imaging) allowing users to freely download, after the publication of this work.

## 2. Background Theory
### 2.1 A new general constraint for phase retrieval of noisy diffraction data: smoothness outside the support region

In CDI, the Fourier magnitudes of an oversampled diffraction pattern, given an oversampling ratio larger than or equal to 2, can in principle be used to retrieve a set of phases that encodes an object (Miao et al., 1998). To recover said phases, iterative algorithms use the Fourier magnitudes as a constraint in reciprocal space and a support in real space. Additional constraints in real space can facilitate the phase retrieval process, but general constraints are difficult to implement and often require prior knowledge of the structure being reconstructed or assumptions about the data being collected. In practice, recovery of an accurate set of phases from oversampled diffraction data in the presence of experimental noise remains a challenge.

Here, we address conditions under which experimental Poisson noise is a predominant component of the noise profile of oversampled diffraction data. In such as case, the high frequency signal is corrupted by high noise (Fig. 1a), which destabilizes the phase retrieval process. We note that the area outside the support can be further exploited to facilitate the faithful recovery of phases that satisfy the Fourier magnitudes. As required by the oversampling condition, the region outside the support is assumed to be zero in ideal cases, but in the presence of noise, it reflects the character of the noise profile of the diffraction intensity. The OSS algorithm presented here applies a general constraint to this region. Namely, OSS forces a smooth density profile onto the region outside the support by means of a convolution with a proper filter at different stage of the iterative process. This is equivalent to the application of a tunable spatial frequency filter (Fig. 1b) to the region outside the support. This frequency filter weighs the contribution of low frequency information in this region more heavily than that of high frequency information, which suffers from a greater degree of corruption from experimental noise. As the total electron density in real space is determined by the centro-pixel value in reciprocal space, applying the smoothness constraint outside the support can reduce the oscillation of the electron density inside the support. By properly choosing spatial frequency filters at different stages of the iteration process, OSS finds a balance between HIO and ER to search for a global minimum, while reducing oscillations in the reconstruction. Furthermore, because the smoothness constraint is applied only to a region outside the support, the spatial resolution and fine features are retained in the reconstruction.

### 2.2 The OSS Framework

In order to search for a global minimum in solution space, OSS starts with 100 independent runs with different phase sets as initial input. With powerful computer clusters, more independent runs can, in principle, be performed to search for larger solution space. Each run iterates back and forth between real and reciprocal space with a total of 2000 iteration. The transition from the j$^{th}$ to the (j+1)$^{th}$ iteration in each run consists of the following steps.

i) Apply the Fourier transform to the input image, $\rho_j(\vec{x})$, and obtain a Fourier pattern, $F_j(\vec{k})$.

ii) Generate a new Fourier pattern by replacing the Fourier magnitudes with the measured ones,
$$F_j^{'}(\vec{k}) = |F_e(\vec{k})| \cdot F_j(\vec{k})/|F_j(\vec{k})| \quad , \quad (1)$$
where $|F_e(\vec{k})|$ represents the experimental Fourier magnitudes.

iii) Calculate a new image, $\rho_j^{'}(\vec{x})$, by applying the inverse Fourier transform to the new Fourier pattern, $F_j^{'}(\vec{k})$.

iv) Revise the image based on the HIO equation (Fienup, 1982),
$$\rho_j^{''}(\vec{x}) = \begin{cases} \rho_j^{'}(\vec{x}) & | \quad (\vec{x} \in S) \cap (\rho_j^{'}(\vec{x}) \geq 0) \\ \rho_j(\vec{x}) - \beta \cdot \rho_j^{'}(\vec{x}) & | \quad (\vec{x} \notin S) \cup (\rho_j^{'}(\vec{x}) < 0) \end{cases} , \quad (2)$$
where $S$ represents a finite support and $\beta$ is a parameter between 0.5 and 1.

v) Calculate the image for the (j+1)$^{th}$ iteration,
$$\rho_{j+1}(\vec{x}) = \begin{cases} \rho_j^{''}(\vec{x}) & | \quad \vec{x} \in S \\ \wp^{-1}[F_j^{''}(\vec{k}) \cdot W(\vec{k})] & | \quad \vec{x} \notin S \end{cases} , \quad (3)$$
where $\wp^{-1}$ is the inverse Fourier transform, $F_j^{''}(\vec{k})$ is the Fourier transform of $\rho_j^{''}(\vec{x})$ and $W(\vec{k})$ is a normalized Gaussian function in Fourier space, defined as
$$W(\vec{k}) = e^{-\frac{1}{2}\left(\frac{\vec{k}}{\alpha}\right)^2} . \quad (4)$$

By modifying parameter $\alpha$, the width of the Gaussian filter can be tuned to reduce the influence of high frequency information in the region outside the support. It is important to note that a Gaussian function $W(\vec{k})$ is used here, but other filter functions can be implemented to suit the needs of particular data. In its present implementation, OSS employs a 10-step function for $\alpha$, shown in Figure 1b. $\alpha$ is linearly changed from on the order of N (step 1) to 1/N (step 10), where N is the array size in that dimension. In step 1, the filter allows nearly all high frequency information to persist outside the support region, exhibiting behavior similar to that of the HIO framework. In step 10, the filter heavily suppresses high frequency information outside the support region, exhibiting behavior similar to that of the ER framework. Each step consists of 200 iterations and the best set of phases with the smallest $R_F$ is passed on as the initial input for the next step. $R_F$, is calculated by,
$$R_F = \frac{\sum_{\vec{k}} \left\| F_e(\vec{k}) \right| - \gamma |F_{j+1}(\vec{k})| \right\|}{\sum_{\vec{k}} |F_e(\vec{k})|} \quad (5)$$

Where $\gamma$ is a scaling factor and $F_{j+1}(\vec{k})$ is the Fourier transform of $\rho_{j+1}(\vec{x})$.

By repeating this iterative process, the algorithm terminates the run after reaching 2000 iterations. The reconstructions of 100 independent runs are compared and the one with the smallest $R_F$ is chosen as the final reconstruction.

## 3. Results
### 3.1 Reconstruction of simulated noisy diffraction patterns

To characterize the effects of different noise levels on the reliability of the phase retrieval process and the accuracy of the retrieved phases, we perform a quantitative comparison among HIO, ER-HIO, NR-HIO and OSS. We first simulate an oversampled diffraction pattern from the Lena model (Fig. 2c). Poisson noise is added to the diffraction intensity with $R_{noise}$ ranging from 5-25%, defined as

$$R_{noise} = \frac{\sum_{\vec{k}} \left\| F_{noise\ free}(\vec{k}) \right| - \left| F_{noise}(\vec{k}) \right\|}{\sum_{\vec{k}} \left| F_{noise\ free}(\vec{k}) \right|} \qquad (6)$$

where $\left| F_{noise\ free}(\vec{k}) \right|$ represents the noise free Fourier magnitudes and $\left| F_{noise}(\vec{k}) \right|$ the Fourier magnitudes with Poisson noise. Figs. 2a and b show the noise free and the noisy Fourier magnitudes ($R_{noise}$ = 15%). Using the same initial sets of random phases and a loose rectangular support, we performed 100 independent runs for each of the four algorithms. Figs. 2d-g show the final reconstructions by HIO, ER-HIO, NR-HIO and OSS, respectively. Visually, the OSS reconstruction is most consistent with the model. Fig. 3a shows the R-factor ($R_F$) as a function of the noise levels for the four algorithms. Although ER-HIO consistently shows a small $R_F$, ER-HIO does not produce the best reconstructions. As ER sets the electron density outside a support to zero in each iteration, the calculation of $R_F$ is dominant by the low spatial frequency of the Fourier magnitudes. A more rigorous method to quantify the reconstructions is to compare them with the model ($R_{real}$), defined as,

$$R_{real} = \frac{\sum_{\vec{x}} \left| \rho_{recon}(\vec{x}) - \rho_{model}(\vec{x}) \right|}{\sum_{\vec{x}} \left| \rho_{model}(\vec{x}) \right|} \qquad , \qquad (7)$$

where $\rho_{recon}(\vec{x})$ represents the final reconstruction by each algorithm and $\rho_{model}(\vec{x})$ the model structure. Fig. 3b shows $R_{real}$ as a function of the noise levels. These results indicate that OSS produces consistently better reconstructions than HIO, ER-HIO and NR-HIO at all noise levels. Next, we performed phase retrieval of a simulated biological vesicle (Fig. 4c). Figs. 4a and b show a noise-free and a noisy diffraction pattern ($R_{noise}$ = 20%), respectively. The final reconstructions of the noisy diffraction pattern by HIO, ER-HIO, NR-HIO and OSS are shown in Figs. 4d-g. From these, it is evident that the OSS algorithm produces the best reconstructions among the four algorithms. Figs. 5a and b show $R_F$ and $R_{real}$ as a function of the noise levels. These simulation results further confirm that OSS produces the most faithful reconstructions at all noise levels among the four algorithms.

### 3.2 Reconstruction of an experimental X-ray diffraction pattern from a Schizosaccharomyces pombe yeast spore cell

To demonstrate the applicability of OSS to experimental data, we performed phase retrieval of an X-ray diffraction pattern measured from a fixed *S. pombe* yeast spore cell (Jiang *et al.*, 2010). Fig. 6a shows the diffraction pattern collected by using 5 keV X-rays from an undulator beamline at SPring-8. A missing center in the diffraction pattern is confined within the centro-speckle, allowing for direct phase retrieval (Miao *et al.*, 2005). By using a loose rectangular support, we perform phase retrieval of the diffraction pattern by using the four algorithms. As a measure of consistency, each phase retrieval algorithm is implemented with five independent trials, each consisting of 100 runs with different initial phase sets. For each of the five trials, the reconstruction with the smallest $R_F$ is chosen as representative of that trial. The reconstructions from the five trials are then compared, and their mean and variance are used as a measure of consistency. Figs. 6d-j show the mean and variance of the five independent trials obtained by HIO, ER-HIO, NR-HIO and OSS, respectively. Visually, OSS produces the most consistent reconstructions. The average $R_F$ and the consistency measure of the trials are shown in Fig. 6b. Although ER-HIO consistently has a small $R_F$ due to the bias towards the low spatial frequency data, the reconstructions obtained by ER-HIO and HIO are more variable from trial to trial than NR-HIO and OSS. Among the four algorithms, OSS produces the most consistent reconstructions with consistency of 96.4%. Furthermore, $R_F$ of OSS is smaller than those of HIO and NR-HIO. This further highlights OSS as a reliable phase retrieval algorithm for the reconstruction of biological specimens from noisy experimental data.

## 4. Discussion

Phase retrieval of oversampled diffraction patterns is fundamentally limited by experimental noise. It remains a challenge to perform consistent phase retrieval of weakly scattering objects such as biological specimens from noisy experimental data. Here we develop the OSS algorithm by implementing a general smoothness constraint upon the region outside the support, which in principle should be zero but in practice reflects the character of the noise profile. We demonstrate that OSS achieves consistent and reliable reconstructions in the presence of experimental noise, conditions in which other phase retrieval algorithms, such as HIO, are more susceptible to corruption by noise.

We expect OSS to improve the consistency and accuracy of phase retrieval efforts from noisy diffraction patterns. The demand for reliable phase retrieval algorithms such as OSS is increasing, given that the imaging of weakly scattering objects, in particular biological specimens, is becoming more popular (Jiang *et al.*, 2010; Miao *et al.*, 2003; Shapiro *et al.*, 2005; Song *et al.*, 2008; Nishino *et al.*, 2009; Huang *et al.*, 2009; Lima *et al.*, 2009; Nelson *et al.*, 2010), and since experimental noise generally limits these applications. Additionally, with the emergence of X-ray free electron lasers (X-FELs), more attention to the treatment of noise is required, given that the diffraction-before-destruction scheme significantly limits the diffraction signal obtained for single X-FEL pulses (Chapman *et al.*, 2006b; Seibert *et al.*, 2011; Mancuso *et al.*, 2010; Schlichting and Miao, 2012), and pose challenges to routinely used phase retrieval algorithms.

## 5. Conclusions

In conclusion, we present here a new phase retrieval framework, termed OSS, which exploits the use of a new general constraint applied to the region outside the support in the iterative process. The constraints implemented by OSS achieve more reliable and faithful reconstructions of noisy diffraction patterns than HIO, ER-HIO and NR-HIO. We demonstrate its validity by using both simulated data with different noise levels, and an experimental data set obtained from a biological cell. We anticipate that OSS will find application in coherent diffraction imaging of a wide range of samples with synchrotron radiation (Miao *et al.*, 1999; Robinson and Harder, 2009; Chapman and Nugent, 2010; Miao *et al.*, 2012; Miao *et al.*, 2002; Pfeifer *et al.*, 2006; Chapman *et al.*, 2006), X-FELs (Chapman *et al.*, 2006b; Seibert *et al.*, 2011; Mancuso *et al.*, 2010; Schlichting and Miao, 2012), high harmonic generation (Sandberg *et al.*, 2007; Sandberg *et al.*, 2008; Ravasio *et al.*, 2009; Seaberg *et al.*, 2011) as well as other fields (Zuo *et al.*, 2003; Dronyak *et al.*, 2009; De Caro et al., 2010; Scott *et al.*, 2012; Bertolotti *et al.*, 2012).

**Acknowledgements**

This work is partially supported by the National Institutes of Health (grant # GM081409-01A1). J.A.R. thanks ??.

**References**

Abbey, B., *et al.* (2008) *Nat. Phys.* **4**, 394-398.

Bertolotti, J., *et al.* (2012) *Nature*. **491**, 232-234.

Chapman, H.N., & Nugent, K.A. (2010) *Nat. Photo.* **4**, 833-839.

Chapman, H.N., *et al.* (2006) *J. Opt. Soc. Am. A.* **23**, 1179-1200.

Chapman, H.N., *et al.* (2006b) *Nat. Phys.* 2, 839-843.

Chen, *et al.* (2007) *Phys. Rev. B.* **76**, 064113.

De Caro, L., *et al.* (2010) *Nat. Nanotech.* **5**, 360-365.

Dronyak, R., *et al.* (2009) *App. Phys. Lett.* **95**, 111908.

Elser, V. (2003) *Acta Cryst. A.* **59**, 201-209.

Fienup, J.R. (1978) *Opt. Let.* **3**, 27-29.

Fienup, J.R. (1982) *Appl. Opt.* **21**, 2758-2769.

Giewekemeyer, K., *et al.* (2010) *Proc. Natl. Acad. Sci. USA*. **107**, 529-534.

Huang, X., *et al.* (2009) *Phys. Rev. Lett.* **103**, 198101.

Jiang, H., *et al.* (2010) *Proc. Natl. Acad. Sci. USA* **107**, 11234-11239.

Lima, E., *et al.* (2009) *Phys. Rev. Lett.* **103**, 198102.

Luke, D.R. (2005) *Inv. Prob.* **21**, 37.

Mancuso, A.P., *et al.* (2010) *New J. Phys.* **12**, 035003.

Marathe, S., *et al.* (2010) *Opt. Exp.* **18**, 7253.

Marchesini, S., *et al.* (2003) *Phys Rev. B.* **68**,140101.

Marchesini, S. (2007) *Rev. Sci. Inst.* **78**, 011301.

Martin, A.V., *et al.* (2012) *Opt. Ex.* **20**, 16650-16661.

Miao, J., Sayre, D., & Chapman, H.N. (1998) *J. Opt. Soc. Am. A.* **15**, 1662-1669.

Miao, *et al.* (1999) *Nature*. **400**, 342-344.

Miao, *et al.* (2002) *Phys. Rev. Lett.* **89**, 088303.

Miao, J., *et al.* (2003) *Proc. Natl. Acad. Sci. USA*. **100**, 110-112.

Miao, J., *et al.* (2005) *Phys. Rev. Lett.* **95**, 085503.

Miao, J., Sandberg, R.L., & Song, C. (2012) *IEEE J. Sel. Top. Quant.* **18**, 399-410.

Nelson, J., *et al.* (2010) *Proc. Nat. Acad. Sci. USA*. **107**, 7235-7239.

Newton, M.C., *et al.* (2012) *Nat. Mater.* **9**, 279-279.

Nishino, Y., *et al.* (2009) *Phys. Rev. Lett.* **102**, 018101.
Pfeifer, M.A., et al. (2006) *Nature.* **442**, 63-66.
Raines, K.S., et al. (2010) *Nature.* **463**, 214-217.
Ravasio, A., *et al.* (2009) *Phys. Rev. Lett.* **103**, 028104.
Robinson, I., & Harder, R. (2009) *Nat. Mater.* **8**, 291-298.
Rodenburg, J.M., *et al.* (2007) *Phys. Rev. Lett.* **98**, 034801.
Roy, S., *et al.* (2011) *Nat. Photo.* **5**, 243.
Sayre, D. (1952) *Acta Cryst.* **5**, 843.
Sandberg, R.L., *et al.* (2007) *Phys. Rev. Lett.* **99**, 098103.
Sandberg, R.L., *et al.* (2008) *Proc. Natl. Acad. Sci. USA.* **105**, 24-27.
Seaberg, M.D., *et al.* (2011) *Opt. Express* **19**, 22470-22479.
Schlichting, I., & Miao, J. (2012) *Curr. Opin. Struct. Biol.* **22**, 613–626.
Scott, M.C., *et al.* (2012) *Nature.* **483**, 444-447.
Seibert, M.M., *et al.* (2011) *Nature.* **470**, 78–81.
Shapiro, D., *et al.* (2005) *Proc. Natl. Acad. Sci. USA.* **102**, 15343-15346.
Song, C., *et al.* (2008) *Phys. Rev. Lett.* **101**, 158101.
Szameit, A., *et al.* (2012) *Nat. Mater.* **11**, 455.
Thibault, P., *et al.* (2008) *Science.* **321**, 379-382.
Williams, G.J., *et al.* (2006) *Phys. Rev. Lett.* **97**, 025506.
Zuo, J.M., *et al.* (2003) *Science.* **300**, 1419-1421.

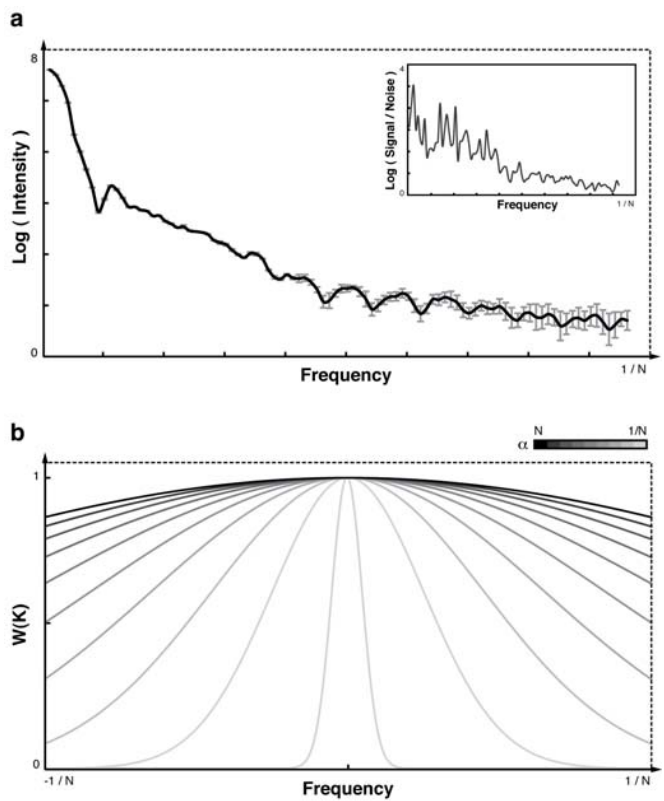

**Figure 1**

(a) A Fourier magnitude profile as a function of the spatial frequency, calculated from a simulated noisy diffraction pattern. The overlaid error bars (gray) indicate the signal variation due to Poisson noise. The inset shows the signal to noise ratio of this Fourier magnitude profile, highlighting the corruption of high frequency information by noise. (b) Line profiles of Gaussian filter functions evaluated for 10 different values of $\alpha$, which is implemented in each of the 10 steps of the OSS algorithm. $\alpha$ is linearly changed from on the order of N in step 1 to 1/N in step 10.

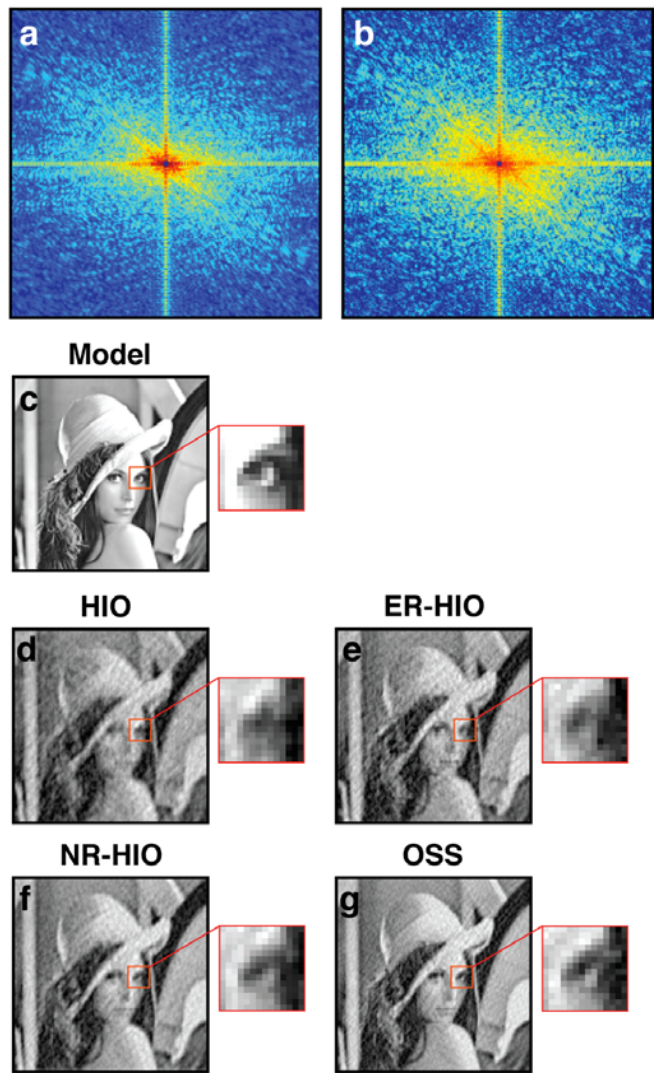

**Figure 2**

(a) Noise-free oversampled diffraction pattern of a Lena model. (b) Oversampled diffraction pattern of the Lena model with Poisson noise ($R_{noise} = 15\%$). (c) Lena model with an inset showing fine features. Final reconstructions of the Lena model obtained from the noisy diffraction pattern shown in (b) by using (d) HIO, (e) ER-HIO, (f) NR-HIO and (g) OSS.

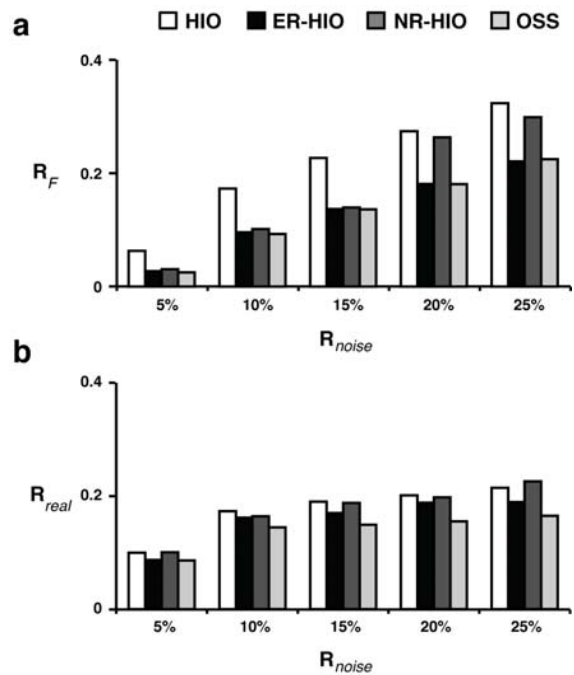

**Figure 3**

(a) R-factor ($R_F$) as a function of the noise levels for the reconstruction of the Lena model. Although ER-HIO has the smallest $R_F$, this does not mean that ER-HIO produces the best reconstruction. As ER sets the electron density outside a support to zero in each iteration, the calculation of $R_F$ is dominant by the low spatial frequency of the Fourier magnitudes. (b) $R_{real}$ (the difference between the final reconstruction and the Lena model) as a function of the noise levels. These simulation results indicate that OSS produces the best reconstructions among the four algorithms at all noise levels.

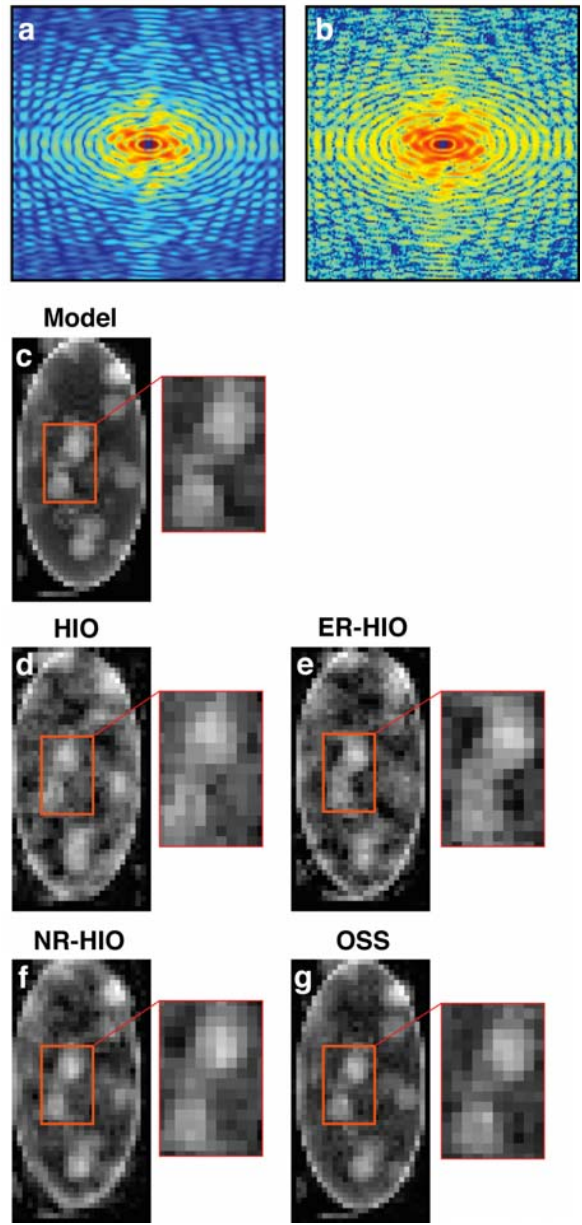

**Figure 4**

(a) Noise-free oversampled diffraction pattern of a biological vesicle model. (b) Oversampled diffraction pattern of the biological vesicle with Poisson noise ($R_{noise}$ = 20%). (c) The biological vesicle model and some fine features (inset). Final reconstructions of the biological vesicle obtained from the noisy diffraction pattern shown in (b) by using (d) HIO, (e) ER-HIO, (f) NR-HIO and (g) OSS.

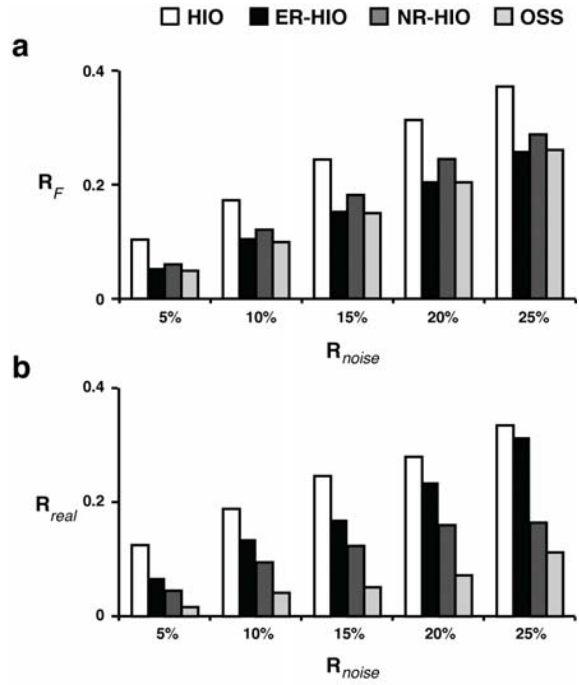

**Figure 5**

(a) R-factor ($R_F$) as a function of the noise levels for the reconstruction of the biological vesicle. (b) $R_{real}$ as a function of the noise levels for the reconstruction of the biological vesicle. These simulation results further confirm that OSS produces the most faithful reconstructions among the four algorithms at all noise levels.

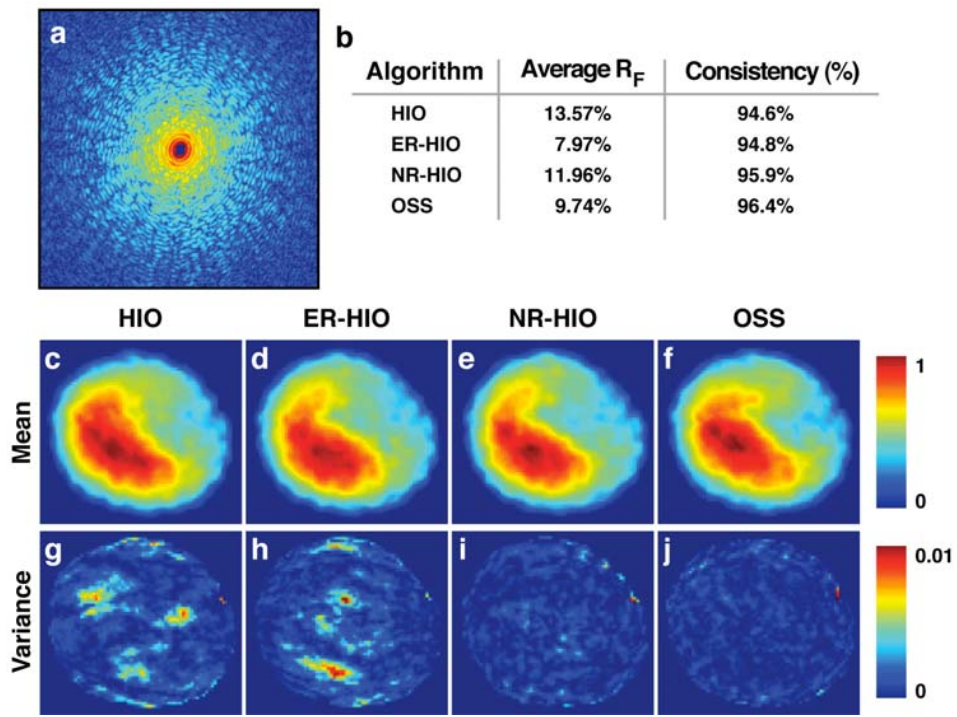

**Figure 6**

(a) Oversampled diffraction pattern measured from a fixed *S. pombe* yeast spore cell using 5 keV X-rays. (b) Average $R_F$ and the consistency of the five independent trials by using the four algorithms. The average reconstruction (mean) of the five independent trials obtained by (c) HIO, (d) ER-HIO, (e) NR-HIO and (f) OSS. Each trial consists of 100 independent runs with different initial phase sets. The variance of the five independent trials obtained by (g) HIO, (h) ER-HIO, (i) NR-HIO and (j) OSS.